\documentclass[aps,preprint,amsmath]{revtex4-1}

\usepackage{graphicx}

\begin{document}


\title{Bifurcation analysis of the transition of dune shape under unidirectional wind}


\author{Hirofumi Niiya}
\author{Akinori Awazu}
\author{Hiraku Nishimori}
\affiliation{Department of Mathematical and Life Sciences, Hiroshima University, Higashihiroshima, Hiroshima 739-8526, Japan}



\date{\today}

\begin{abstract}
A bifurcation analysis of dune shape transition is made.
By use of a reduced model of dune morphodynamics, {\it dune skeleton model}, we elucidate the transition mechanism between different shapes of dunes under unidirectional wind.
It was found that the decrease in the total amount of sand in the system and/or the lateral sand flow shifts the stable state from a straight transverse dune to wavy transverse dune through a pitchfork bifurcation.
A further decrease causes wavy transverse dunes to shift into barchans through a Hopf bifurcation.
These bifurcation structures reveal the transition mechanism of dune shapes under unidirectional wind.
\end{abstract}

\pacs{05.45.-a, 45.70.Qj, 45.70.Vn, 92.40.Gc}

\maketitle

Sand dunes, considered as the largest granular objects on Earth, have several distinct shapes, for instance, barchan, transverse, linear, star-shaped, dome-shaped, and parabolic dunes
\cite{mckee1979introduction,cooke1993desert}.
These shapes are governed by two dominant factors; the steadiness of the wind direction and the amount of available sand
\cite{livingstone1996aeolian}.
For example, unidirectional wind generates barchans and transverse dunes.
The former, crescent-shaped isolated dunes, are formed in dune fields with small amounts of available sand, whereas the latter, extending perpendicular to the wind direction, are formed in dune fields with larger amounts of available sand than barchan-rich fields.
Recent dune studies have reflected a significant progress in the quantitative analysis of dune morphodynamics.
For instance, rescaled water tank experiments have successfully been conducted to form distinct dune shapes under controlled conditions
\cite{hersen2002relevant,endo2005barchan,groh2008barchan}.
Further, computational models have been used to reproduce various shapes of dunes
\cite{nishimori1993formation,werner1995eolian,nishimori1998simple,duran2005breeding,zhang2010morphodynamics,katsuki2011cellular}.
However, a theoretical methodology to explain the basic mechanism behind dune shape formation beyond a mere numerical reproduction of the shapes is yet to be established except for a few studies like simple discussion by Parteli et al.
\cite{parteli2011transverse}.
To address this lack of analytical methodology, Niiya {\it et al}. proposed the {\it Dune skeleton model} (hereafter, {\it DS model}) consisting of coupled ordinary differential equations, each of which represents the dynamics of two-dimensional cross sections (hereafter, 2D-CSs) of a three-dimensional dune
\cite{niiya20103d,niiya2011erratum}.
So far, the model has successfully reproduced 3 typical shapes of dunes--straight transverse dune, wavy transverse dune, and barchan--depending on the amount of available sand and wind strength.
In this study, using bifurcation analysis, we investigate the shape transition of dunes with the change in relevant parameters.

\begin{figure}[t]
\begin{center}
\includegraphics[width = 3.0 in]{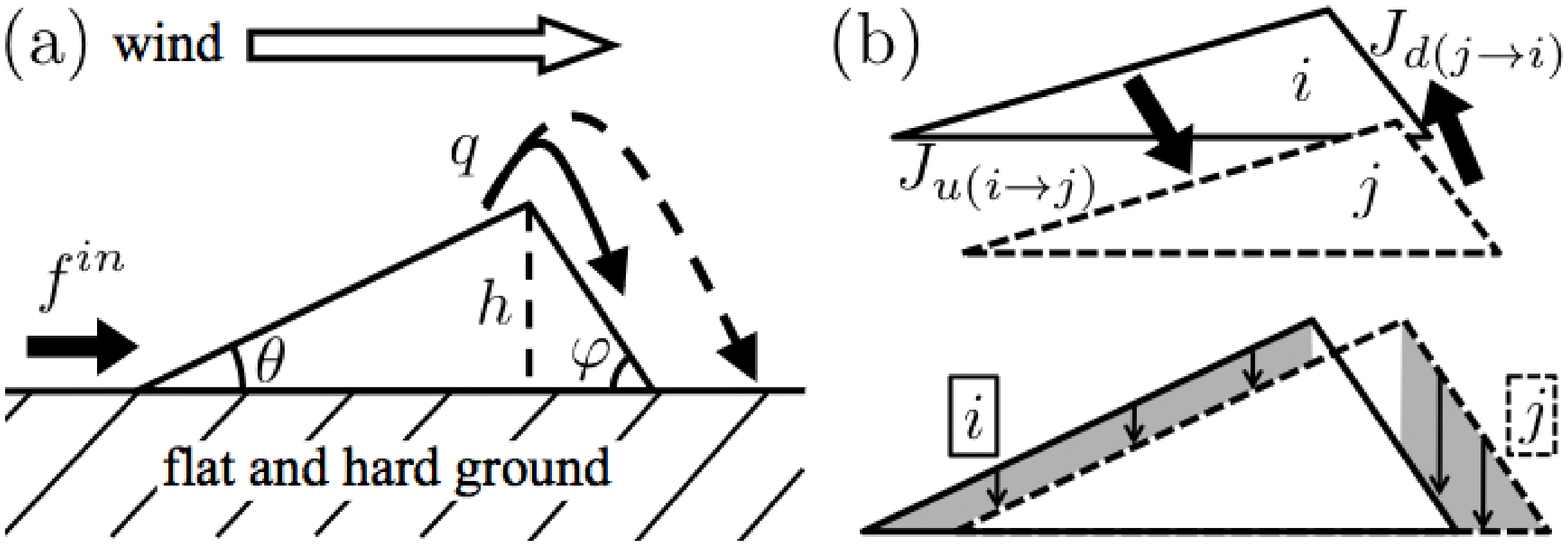}
\caption{
(a) Intra-2D-CS sand flow.
(b) Inter-2D-CS sand flow.
}
\label{fig:sand-flow}
\end{center}
\end{figure}

The {\it DS model} covers the formation processes of barchans and transverse dunes, both of which are generated under unidirectional steady wind.
This model is roughly based on two considerations.
First, a lateral array of 2D-CSs is set perpendicular to the wind direction with constant intervals.
Second, a combination of two forms of sand movement--the intra-sand movement within each 2D-CS and the inter-sand movement between neighboring 2D-CSs--is considered in governing the macroscopic morphodynamics of dunes.
Considering the observation that 2D-CSs of barchans and transverse dunes very roughly show a scale invariant triangular shape,
we assume that the angles of their upwind and downwind slopes ($\theta$ and $\varphi$, respectively) are maintained constant through their time evolution, irrespective of their size
(Fig. \ref{fig:sand-flow}(a)).
We also assume that dunes are located on a flat and hard ground.
From these assumptions, the horizontal (i.e., wind directional) position and the height of each 2D-CS is uniquely determined if only the coordinate ($x,h$) of its crest is given.
Moreover, empirical geometrical constants A, B, and C are set to 1/10, 4/5, and 1/5, respectively, reflecting the typical 2D-CS profiles of real barchans and transversed dunes.
As mentioned above, sand flow is classified into two forms:
(a) the intra-2D-CS flow and (b) the inter-2D-CS flow.
The intra-2D-CS flow along the upwind slope is uniquely determined if the overcrest sand flux $q$ and the incoming sand flux from the windward ground $f^{in}$ are given
(Fig. \ref{fig:sand-flow}(a)).
The overcrest sand flux $q$ deposits along the downwind slope or escapes to the leeward inter-dune ground.
The deposition ratio $T_E$ in the downwind slope, termed {\it sand trapping efficiency} \cite{cooke1993desert}, is basically an increase function of height with upper limit $T_E=1$.
Considering a real-data-based estimation of $T_E$ by Momiji et al. \cite{momiji2000relations},
we roughly assumed the specific form of $T_E$ as
\begin{eqnarray*}
T_E(h)=\frac{h}{1.0 + h},
\end{eqnarray*}
here $h$ is the non-dimensionalized height of dune with the physical unit of meter
\cite{ste}.
The quantity $q$ reflects the over-dune wind strength; here, we assume $q$ to be a constant, independent of the 2D-CS height.
In addition, all of the incoming sand flux from the windward ground, $f^{in}$, deposits on the upwind slope.


The inter-2D-CS flow $J_{u(i\to j)}/J_{d(j\to i)}$ occurs only between the upwind/downwind slopes of neighboring 2D-CSs, $i$ and $j$.
The flux is roughly considered as lateral diffusion depending on the height difference, though the consideration of the overlap length of slopes causes nonlinearity in this inter-2D-CS flux.
The specific forms of $J_{u(i\to j)}$ and $J_{d(j\to i)}$ are described as
%
{\small
\begin{subequations}
\begin{eqnarray}
\label{eq:inter-up}
J_{u(i\to j)}=
\left\{
\begin{array}{lr}
 \cfrac{D_u{\rm B}}{{\rm 2A}\Delta w^2}\left\{h_i^2-\left[h_j-\cfrac{\rm A}{\rm B}(x_j-x_i)\right]^2\right\} & x_j-x_i > 0\\
 \cfrac{D_u{\rm B}}{{\rm 2A}\Delta w^2}\left\{\left[h_i+\cfrac{\rm A}{\rm B}(x_j-x_i)\right]^2-h_j^2\right\} & x_j-x_i \le 0
\end{array}
\right.\\
\label{eq:inter-down}
J_{d(j\to i)}=
\left\{
\begin{array}{lr}
 \cfrac{D_d{\rm C}}{{\rm 2A}\Delta w^2}\left\{h_j^2-\left[h_i-\cfrac{\rm A}{\rm C}(x_j-x_i)\right]^2\right\} & x_j-x_i > 0\\
 \cfrac{D_d{\rm C}}{{\rm 2A}\Delta w^2}\left\{\left[h_j+\cfrac{\rm A}{\rm C}(x_j-x_i)\right]^2-h_i^2\right\} & x_j-x_i \le 0,
\end{array}
\right.
\end{eqnarray}
\end{subequations}
}
where $\Delta w$ is the lateral interval between neighboring 2D-CSs and is set as $\Delta w=1$ hereafter
\cite{unit}.
These quantities correspond to the grey areas in Fig. \ref{fig:sand-flow}(b) multiplied by the diffusion coefficients.
Here, the upwind and downwind diffusion coefficients ($D_u$ and $D_d$, respectively) control the amount of inter-2D-CS sand flow on respective sides of the slopes and reflect the over-dune wind strength.
With consideration of the above intra- and inter-sand flows, the dynamics of the coordinate $(x_i,h_i)$ of 2D-CS's crest is given as a system of coupled ordinary differential equations:
%
%
{\small
\begin{subequations}
\begin{align}
\label{eq:dsma}
\frac{dx_i}{dt}=&\frac{1}{h_i}\left[q({\rm B}T_E(h_i)+{\rm C})+\sum_{j=i\pm1}({\rm B}J_{d(j\to i)}+{\rm C}J_{u(i\to j)})-{\rm C}f_i^{in}\right],\\
\label{eq:dsmb}
\frac{dh_i}{dt}=&\frac{\rm A}{h_i}\left[q(T_E(h_i)-1)+\sum_{j=i\pm 1}(J_{d(j\to i)}-J_{u(i\to j)})+f_i^{in}\right].
\end{align}
\end{subequations}
}
We also introduce the annihilation rule of the 2D-CSs at the lateral edges of a dune;
This rule is required to simulate the shrinking process of dunes.
This rule is applicable in the cases where $h_i$ decreases to $h_i=0$ or where the overlap of either the upwind or downwind slope between the 2D-CS at the edge and its nearest neighbor vanishes.
In the previous study, numerical simulations of eqs. (\ref{eq:dsma}) and (\ref{eq:dsmb}) have been conducted with $N=1000$ 2D-CSs.
The lateral boundary condition was set to be periodic, whereas the wind directional boundary condition was set such that the total amount of escaped sand is uniformly redistributed on the windward ground.
These simulations showed that three typical shapes of dunes--straight transverse dunes, wavy transverse dunes, and barchans--are formed depending on the combination of the initial average height of 2D-CSs (in other words, the amount of sand), the overcrest sand $q$, and the upwind diffusion coefficient $D_u$.

\begin{figure}[tb]
\begin{center}
\includegraphics[width = 3.0 in]{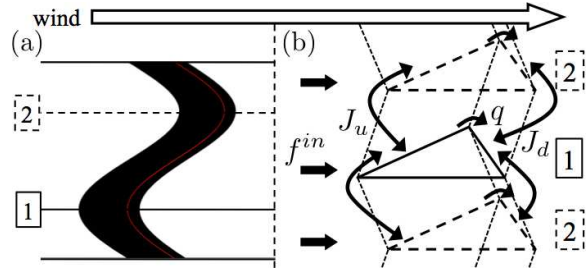}
\caption{
Analysis setup:
(a) 
Two 2D-CSs selected from the transverse dune obtained by simulation:
first and second 2D-CSs are represented by the solid and dashed lines, respectively.
(b) 
System of two 2D-CSs with lateral periodicity.
}
\label{fig:analysis-setup}
\end{center}
\end{figure}

In this study, we elucidate the transition mechanism between different types of dunes.
To do so, we conduct a bifurcation analysis of the reduced {\it DS model}.
The reduced {\it DS model} consists of two 2D-CSs, each of which is assumed to represent a large number of 2D-CSs used in the previous simulation (Fig. \ref{fig:analysis-setup}).
The reduced {\it DS model} is described by four variables $(x_1,h_1,x_2,h_2)$, which are the coordinates of the two 2D-CS's crests.
The lateral boundary condition is set to be periodic, while the wind directional boundary condition is set such that sand escaped from the leeward boundary is redistributed uniformly at the windward boundary.
Because of these boundary conditions, the total amount of sand in the system is preserved.
Therefore, the total area of the 2D-CSs,
\begin{eqnarray}
\label{eq:conserved-sand}
S=\frac{h_1^2+h_2^2}{\rm 2A}\equiv \frac{H_0^2}{\rm A},
\end{eqnarray}
is maintained constant unless the annihilation of a 2D-CS occurs, and we do not treat the annihilation process in this study.
The relation eq. (\ref{eq:conserved-sand}) is rewritten as $h_2(h_1)=\sqrt{2 H_0^2-h_1^2}$.
In addition, by defining the relative distance $y=x_2-x_1$ between two 2D-CSs, the reduced {\it DS model} is further reduced into a two-variable system.
Moreover, according to the rule for the redistribution of escaping sand, the incoming sand fluxes from windward ground $f_i^{in}$ are expressed as
\begin{eqnarray*}
f_i^{in}=\frac{1}{2}\sum_{j=1}^{2}q(1-T_E(h_j)),\hspace{4 ex}i=1,2
\end{eqnarray*}
where $T_E(h)$ is the deposition ratio in the downwind slope given above.
In addition, symmetric relations of the inter-2D-CS flow, $J_{u(1\to 2)}=-J_{u(2\to 1)}$ and $J_{d(2\to 1)}=-J_{d(1\to 2)}$, are obtained from the definitions of the four variables constituting the two 2D-CSs.
Then, the dynamics of the reduced {\it DS model} are described as follows:
%
%
{\small
\begin{subequations}
\begin{align}
\nonumber
\frac{dy}{dt}=&q{\rm B}\left(\frac{T_E(h_2(h_1))}{h_2(h_1)}-\frac{T_E(h_1)}{h_1}\right)+{\rm C}(q-f_1^{in})\left(\frac{1}{h_2(h_1)}-\frac{1}{h_1}\right)\\
\label{eq:two-eqa}
&-2({\rm B}J_{d(2\to 1)}+{\rm C}J_{u(1\to 2)})\left(\frac{1}{h_2(h_1)}+\frac{1}{h_1}\right),\\
\label{eq:two-eqb}
\frac{dh_1}{dt}=&\frac{\rm A}{h_1}\left[q(T_E(h_1)-1)+2(J_{d(2\to 1)}-J_{u(1\to 2)})+f_1^{in}\right].
\end{align}
\end{subequations}
}
In this simplest form of the reduced {\it DS model}, we vary 4 parameters, $H_0, q, D_u,$ and $D_d$.
$H_0$ controls the total amount of sand in the system;
$q$ is the overcrest sand, which reflects the wind strength in a dune field;
and $D_u$ and $D_d$ are the upwind and downwind diffusion coefficients, respectively, as introduced in eqs. (\ref{eq:inter-up}) and (\ref{eq:inter-down}).
$D_u$, similar to $q$, is influenced by the wind strength, although the correlation between $D_u$ and $q$ is not discussed here.
Now, we conduct the analysis of eqs. (\ref{eq:two-eqa}) and (\ref{eq:two-eqb}).
We obtain two types of fixed points:
a) fixed points with $h_1=h_2$ and $y=0$, referred to as ``straight fixed points'' and denoted as $S_F$,
and b) fixed points with $h_1\ne h_2$ and $y\ne 0$, referred to as ``wavy fixed point" and denoted as $W_F$.
To avoid the cases requiring the annihilation rule of 2D-CS, we consider the inter-2D-CS flux function (eqs. (\ref{eq:inter-up}) and (\ref{eq:inter-down})) and search for fixed points of eqs. (\ref{eq:two-eqa}) and (\ref{eq:two-eqb}) under the limited conditions where the interaction between two 2D-CS is kept non-zero; these conditions are given as
\begin{eqnarray}
\label{eq:range}
0<h_1<\sqrt{2}H_0,\hspace{1 ex} -\frac{\rm C}{\rm A}\sqrt{2}H_0 < y < \frac{\rm C}{\rm A}\sqrt{2}H_0.
\end{eqnarray}
$S_F$ always exists within these ranges; that is, $h_1=h_2=H_0$ and $y=0$ serve as a fixed point of eqs. (\ref{eq:two-eqa}) and (\ref{eq:two-eqb}), independent of control parameters $q, D_u,$ and $D_d$.
However, these control parameters largely affect the stability of $S_F$ and the existence of $W_F$.
Here, in order to relate the fixed points $S_F$ and $W_F$ to dune shapes obtained by our previous simulation of the full {\it DS model} with $N=1000$ 2D-CSs as well as to those in the desert field,
we check the existence of $W_F$ and the stability of $S_F$ and $W_F$.
\begin{description}
\item[I)]
If $S_F$ is stable, we consider this state as corresponding to the stable straight transverse dunes.
The phase of the system in this state is referred to as the ``ST phase.''
\item[II)]
If $S_F$ becomes unstable and, in turn, $W_F$s emerge as stable fixed points, we consider this transition as corresponding to that from the straight transverse dunes to the stable wavy transverse dunes.
The phase of the system in this state is referred to as the ``WT phase.''
\item[III)]
If $W_F$ becomes unstable with $S_F$ already unstable,
we consider this transition as corresponding to that from wavy transverse dunes to barchans.
The phase of the system in this state is referred to as the ``B phase.''
\end{description}
In the B phase, no stable fixed point is obtained.
Correspondingly, a single barchan is not stable in the {\it DS model} simulation and gradually shrinks with time.
Note that the laterally periodic boundary condition of the present model should be reconsidered to correctly describe the dynamics of barchans.
\begin{figure}[t]
\begin{center}
\includegraphics[width= 3.0 in]{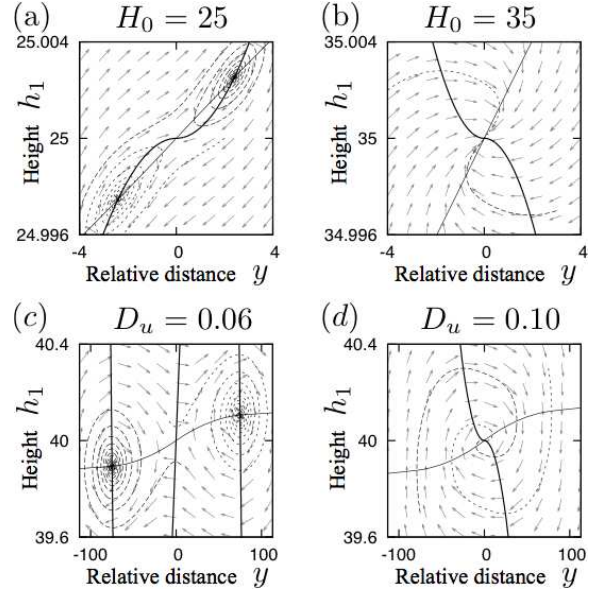}
\caption{
Nullclines, solution trajectories, and vector fields of eq. (\ref{eq:two-eqa}) and (\ref{eq:two-eqb}) for (a) $H_0=25$, (b) $H_0=35$, (c) $D_u=0.06$, and (d) $D_u=0.10$.
The thin and thick solid lines represent $y$-nullcline and $h_1$-nullcline, respectively.
Both short and longer dashed lines represent the solution trajectories starting from the different initial values.
The vectors are normalized in magnitude while maintaining the direction.
The unvarying parameters are set as $q=0.5$ and $D_d=0.1$ in all figures.
Other parameters are set as ((a), (b)) $D_u=0.1$ and ((c), (d)) $H_0=40$.
}
\label{fig:nullcline-sand}
\end{center}
\end{figure}
Now, we study the phase change transition between ST and WT phases and that between WT and B phases using the bifurcation analysis.
First, we consider the nullclines of eqs. (\ref{eq:two-eqa}) and (\ref{eq:two-eqb}) as a two-dimensional dynamical system,
where a pair of control parameters are fixed as $q=0.5$ and $D_d=0.1$ and the remaining two parameters, $H_0$ and $D_u$, are varied independently.
In Fig. \ref{fig:nullcline-sand}, the nullclines in the $y$-$h_1$ phase space are drawn at typical values of $H_0$ or $D_u$.
They suggest that the decrease in $H_0$ and/or $D_u$ plays a significant role in the emergence of $W_F$.
Here, the results are unchanged even if the two 2D-CSs are exchanged; that is, below a critical value of $H_0$ or $D_u$, two points of $W_F$ simultaneously emerge and are symmetrical around $S_F$ (Fig. \ref{fig:nullcline-sand}(a), (c)).
Next, in order to determine the bifurcation structure and stability of fixed points, we search for fixed points by continuously varying the parameters $H_0$ and $D_u$ within the range satisfying relation eq. (\ref{eq:range}) and conduct the linear stability analysis (Fig. \ref{fig:bifurcation}).
The bifurcation diagrams clearly show the transitions between three phases: ST, WT, and B phases.
Figure \ref{fig:bifurcation}(a) and \ref{fig:bifurcation}(b) shows that the decrease in $H_0$ and/or $D_u$ destabilizes $S_F$ to shift the corresponding phase of the system from ST to WT and further decrease in the same destabilizes $W_F$ to shift the system into B phase.
It should be noted that B phase in Fig. \ref{fig:bifurcation}(b) occurs when $W_F$ goes out of the range of eq. (\ref{eq:range}).
\begin{figure}[t]
\begin{center}
\includegraphics[width= 3.0 in]{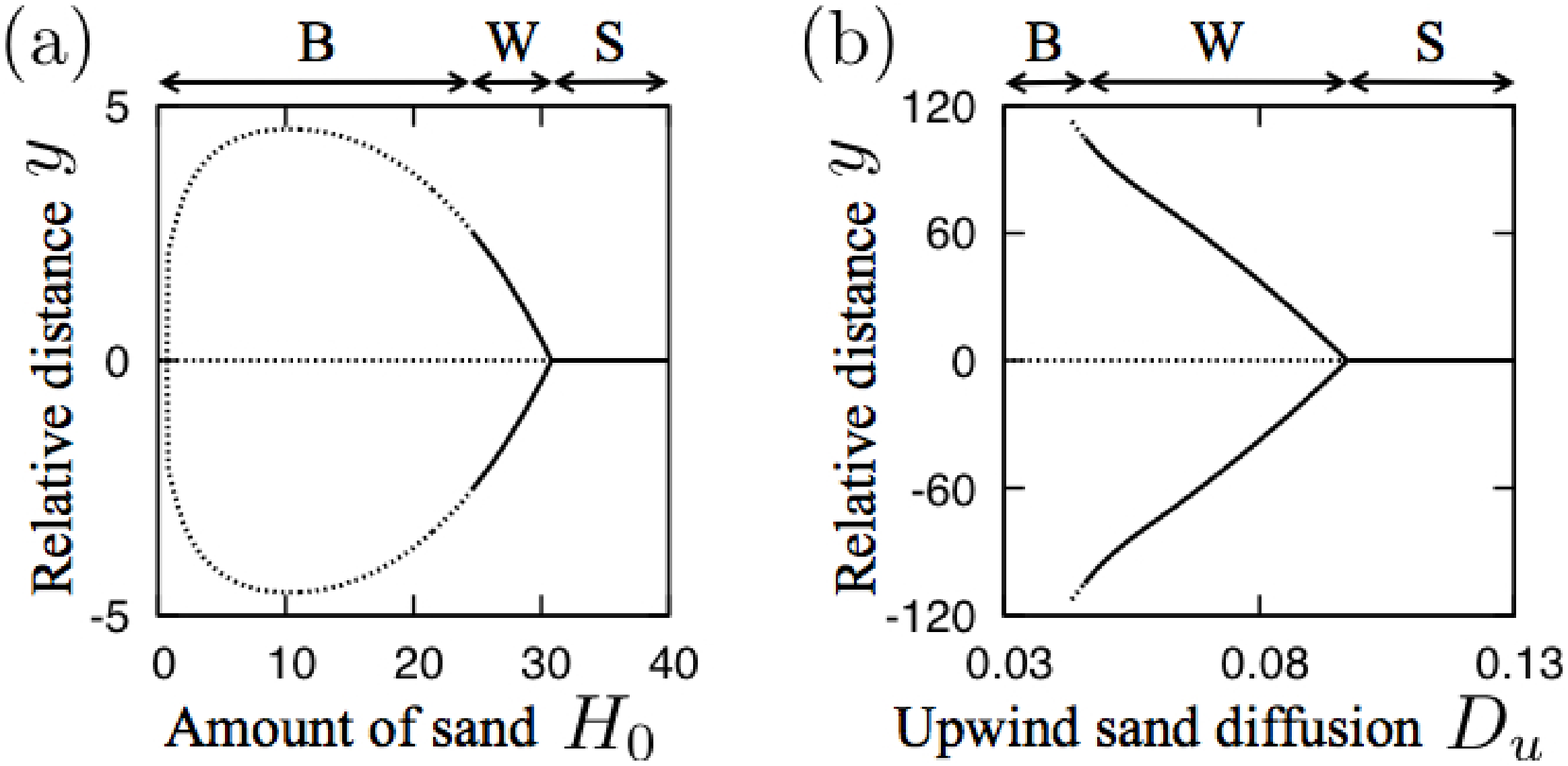}
\caption{
Bifurcation diagram of fixed points obtained by varying one control parameter: (a) $H_0$ and (b) $D_u$.
The solid and dashed lines denote stable and unstable fixed points, respectively.
B, W, and S denote the ranges of B phase, WT phase, and ST phase, corresponding to the dune shapes, respectively.
The unvarying parameters $(q, D_d)$ that are common between (a) and (b) are the same as the values in Fig. \ref{fig:nullcline-sand},
whereas other parameters are set as (a) $D_u=0.1$ and (b) $H_0=40$.
}
\label{fig:bifurcation}
\end{center}
\end{figure}
Moreover, in Fig. \ref{fig:bifurcation}(a) and \ref{fig:bifurcation}(b), we focus on two bifurcation points in the respective figures:
(i) first bifurcation points of switching from ST phase to WT phase and
(ii) second bifurcation points of switching from WT phase to B phase.
The type of bifurcation at the first bifurcation points is the pitchfork bifurcation because the stable $S_F$ becomes unstable and two symmetric stable $W_F$ points arise with the decrease in $H_0$ and $D_u$.
Further decrease in $H_0$ and $D_u$ destabilizes $W_F$, however, it does not cause the emergence of new fixed points.
Beyond these second bifurcation points, the eigenvalues of the Jacobian matrix at $W_F$ contain imaginary parts from the real phase space.
This type of bifurcation at the second bifurcation points is the Hopf bifurcation.
Finally, we classify each of the two sets of parameter spaces $(H_0,q)$ and $(D_u,q)$ into three phases using the bifurcation analysis discussed above.
These phase diagrams qualitatively correspond to our previous numerical simulation results of the {\it DS model}
\cite{niiya20103d,niiya2011erratum}.
Namely, an increase in the amount of sand and inter-2D-CSs flow enhances the stability of the transverse dune, whereas the decrease in intra-2D-CS flow destabilizes its shape to enforce the deformation to a barchan.
Moreover, these shape transition reproduce the change in real dune shapes depending on the amount of available sand.

In this study, we dealt with dune morphodynamics from a dynamical system point of view.
A simple model called the {\it Dune Skeleton model} was further simplified to a two-variable ordinary differential equation named the ``{\it reduced Dune Skeleton model}.''
This {\it reduced DS model} enabled the bifurcation analysis for the transition between different dune shapes observed under a unidirectional wind.
The results qualitatively corresponded to the previous observations of real dunes, water tank experiments, and simulation models.
But the mere careful comparison to recent study which are made under similar but not identical condition is next issue.

This work was partially supported by the 21st Century COE Program, ``Toward a new basic science depth and synthesis.''


\begin{thebibliography}{99}
\bibitem{unit} All non-dimensionalized quantities used in the present study are given with the physical units of the scale of meter and day.
\end{thebibliography}

\end{document}